\documentclass[conference]{IEEEtran}
\IEEEoverridecommandlockouts
\usepackage{cite}
\usepackage{amsmath,amssymb,amsfonts}
\usepackage{graphicx}
\usepackage{textcomp}
\usepackage{algorithm}
\usepackage{pdfpages}
\usepackage{subcaption}
\usepackage{xcolor}
\usepackage{multirow}
\usepackage{comment}
\def\BibTeX{{\rm B\kern-.05em{\sc i\kern-.025em b}\kern-.08em
    T\kern-.1667em\lower.7ex\hbox{E}\kern-.125emX}}
    
\usepackage[noend]{algpseudocode}
\usepackage{graphicx}

\usepackage{adjustbox}

\usepackage{graphicx}
\usepackage{algorithm}

\usepackage{url}

\begin{document}
\title{DeepTaskAPT: Insider APT detection using Task-tree based Deep Learning}
%
%
\author{\IEEEauthorblockN{ Mohammad Mamun}
\IEEEauthorblockA{\textit{Digital Technology} \\
\textit{National Research Council Canada}\\
Fredericton, Canada \\
mohamad.mamun@nrc-cnrc.gc.ca}
\and
\and
\IEEEauthorblockN{Kevin Shi}
\IEEEauthorblockA{\textit{Digital Technology} \\
\textit{National Research Council Canada}\\
Fredericton, Canada \\
kevin.shi@nrc-cnrc.gc.ca}
}

\maketitle
\begin{abstract}

APT, known as Advanced Persistent Threat, is a difficult challenge for cyber defence. These threats make many traditional defences ineffective as the vulnerabilities exploited by these threats are insiders who have access to and are within the network. This paper proposes DeepTaskAPT, a heterogeneous task-tree based deep learning method to construct a baseline model based on sequences of tasks using a Long Short-Term Memory (LSTM) neural network that can be applied across different users to identify anomalous behaviour. Rather than applying the model to sequential log entries directly, as most current approaches do, DeepTaskAPT applies a process tree based task generation method to generate sequential log entries for the deep learning model.


To assess the performance of DeepTaskAPT, we use a recently released synthetic dataset, DARPA Operationally Transparent Computing (OpTC) dataset and a real-world dataset, Los Alamos National Laboratory (LANL) dataset. Both of them are composed of host-based data collected from sensors. Our results show that DeepTaskAPT outperforms similar approaches e.g. DeepLog and the DeepTaskAPT baseline model demonstrate its capability to detect malicious traces in various attack scenarios while having high accuracy and low false-positive rates. To the best of knowledge this is the very first attempt of using recently introduced OpTC dataset for cyber threat detection.  


\end{abstract}

\IEEEpeerreviewmaketitle

\section{Introduction}
\label{sec:introduction}

Cyber defense is a fundamental problem for the digital economy, and it is, at the root, a data synthesis problem. Organizations set up multiple sensors to feed their cyber defenders with concurrent data streams reporting a large and eclectic set of observations regarding resource usage, network communications, application logs, user and host behavior, threat intelligence, and so on. Defenders maintain awareness of ongoing activities, and of malicious activities in particular, by making sense of this complicated dataset. One of these data synthesis tasks is to detect \emph{anomalies} in the data stream, e.g. extraordinary sequences of events, under the assumption that incipient incidents would involve behaviors or actions observed most rarely.


Current log-based methods of anomaly detection can be narrowly grouped into two groups: Approaches based on log event indices and approaches based on log template-semantics \cite{Huang2020}. Log event indices anomaly detection methods i.e. \cite{Lou2010, He2018, Du2017, Huang2020} extracts log events from log messages and converts them into indexes. This approach doesn't attempt in log messages to use semantic knowledge. Thus, unseen log models cannot be handled and can be unreliable. On the other hand, log template semantic based methods \cite{Meng2019, Mamun2016, Zhang2019} treat the log stream model as a natural language sequence and transform log templates to word vectors to train the model. Both of these techniques can be used successfully to detect advanced persistent threat (APT) to the modern enterprise \cite{log2vec2019}.

We focus on understanding advanced persistent threat (APT) attacks against host-based sensor telemetry and develop new tools to combat them.
Host-based sensors are headless software with features similar to antivirus or EDR agents. They capture low-level events regarding process life cycles, network transactions, file operations, and other services of the operating system; and relay all these events to a central repository. Host-based telemetry is a heterogeneous dataset. It is composed of event records of varying schema, whose semantics differ significantly between event types.
These events enable analysts to reconstruct the various threads of activity occurring on the host, particularly that of an APT actor as it deploys and runs persistently on the host to compromise other hosts via intranet and steal sensitive information \cite{Bohara2017}. Many other similar works restrict themselves to a subset of event types, but we hypothesize that the collective semantics of events of various types enables a better modeling of normal event flows \cite{Mamun2019}. In addition, concurrent activity threads, such as web browsing, text redaction, system housekeeping are superposed in a single stream. In the same vein, certain host activities generate long sequences of events of a single type. As normal activity can mix such data phenomena in a combinatorially large number of ways, the development of recurrent neural models is complicated.

Recently, recurrent techniques used in natural language processing \cite{Zhang2016} have been applied to log data analysis, for purposes of system failure diagnosis and root cause analysis. In \cite{Young2019}, a clustering technique is used to detect and forecast device failure through several log entries that are input to the LSTM network. \emph{DeepLog} in \cite{Du2017} has used a generalised method of identification and diagnosis, with tasks isolated from a log file. For each task a working flow model has been created before feeding to the LSTM model. \emph{OCAN} in \cite{PanPan2019}, a semi-supervised anomaly detection model, uses LSTM for fraud detection learning the representations of users from their web activity. \emph{DeepAPT} in \cite{Ishai2017} demonstrates how deep neural network can be used to attribute nation-state APTs using sandbox reports as input. That said, advances based on deep learning can play a role in addressing issues with the aforementioned work, in terms of detection performance, typically due to the novelty of anomalies raised from malicious activity. Learning rich normality representations with a limited amount of data (with labeled anomalies) that generalize to new types of anomalies remains a major challenge in unsupervised/semi-supervised anomaly detection \cite{Aggar2017}. Deep learning methods such as LSTM allow the whole anomaly detection pipeline to be optimised and facilitate learning representation designed for the detection of unknown anomalies \cite{Jie2019,Dan2017}. The methodology embraced in these papers inspires the methods we present here.




In general, a typical APT detection method transforms user operations into sequences that can store information, such as the sequential relationship between log entries, and then uses sequence modelling techniques, such as N-gram to learn from past events and predict the next one. In essence, these methods model user behaviour at the training stage and trigger exceptions as anomalies. However, concurrency is another big challenge in this domain. It is certain that the order of events in a log contains valuable insights and analytical detail, but events log in the host can be generated by many different users, threads or concurrent tasks. Prediction approaches based on continuous logs can suffer a reliability loss to APT detection if this relationship in the log is ignored.






\subsection{Our contribution}

We propose DeepTaskAPT, a deep learning method based on tasks performed by the user, keeping in mind \textit{task-based relationships} in the log to detect APT attacks. DeepTaskAPT comprises three components:  
1) task tree construction. DeepTaskAPT constructs a task tree based on hierarchical relationships between the system processes to integrate relationships between log entries to determine users \textit{concurrent tasks} in the process trees within a host. 2) a baseline model. A Long Short-Term Memory (LSTM) based neural network model that includes all types of host-based events in a users' tasks to vectorize the user's normal activities to allow for a plausible evaluation of their similarity to identify anomalies. This powerful approach is a classifier trained solely on normal usage data without any assumption on the deployed malicious tactics or common malware categories. 3) an anomaly detector, against the baseline model to identify malicious actions effectively. We assess the performance of DeepTaskAPT against the DARPA OpTC \cite{OpTC2020} and LANL \cite{Turcotte2018} synthetic and real-life datasets respectively. They are highly representative of host-based data streams captured through enterprise-grade sensors. Furthermore, DeepTaskAPT can be updated incrementally from new data, new users, as well as the identification of false positives by a human analyst. This adaptive aspect of the model makes it appealing as its training cost can be spread over time, enabling higher mission availability.





\section{Overview}
\label{sec:review}


\subsection{Anomaly detection in host-based telemetry}
\label{sec:contribution}

Semantics-aware anomaly detection methods such as DeepLog in \cite{Du2017} transform user operations into sequences that store information, such as the sequential relationship between log entries. They use sequence modeling techniques, such as $N$-gram decomposition, to predict the next event from history. In essence, these methods model user behaviour at the training stage and trigger exceptions as anomalies.

However, these methods may overlook other relationships. For example, a large number of operations at any given time may imply a data breach and can be detected by the trained LSTM model based on the user's regular behavior \cite{log2vec2019}.  
This surge of operations (sequences ordered by time) may not be related to each other. Clearly, \textit{sequential} relationship among the operations may not be a \textit{logical} relationship. In fact, they might be generated from concurrent tasks. Existing sequence based deep learning methods ignore these relationships. It seems to be a strong assumption that a user's everyday behaviour must be fairly consistent and comparable over time. Recently Log2vec in \cite{log2vec2019} proposes a complex graph embedding based approach to address this problem and shows that existing approaches' performance incompetency to detect APT attacks.          

\begin{figure*}[h]
    
    \hspace*{\fill}   
    \begin{subfigure}{0.74\textwidth}
        \includegraphics[width=\textwidth]{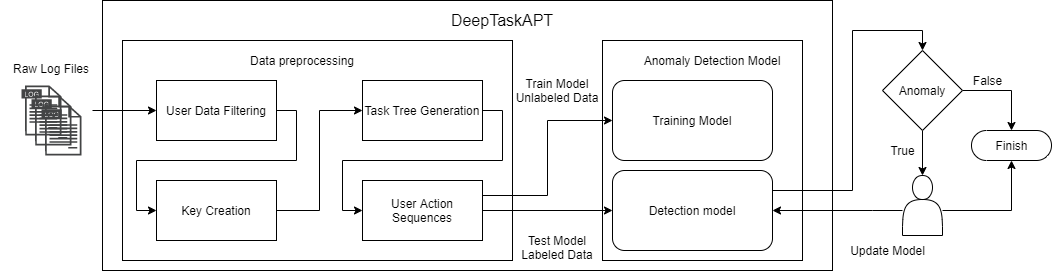}
        \caption{}
        \label{DeepTaskAPT Architecture}
    \end{subfigure}%
    \begin{subfigure}{0.25\textwidth}
        \centering
        \includegraphics[width=4cm,height=5cm,keepaspectratio]{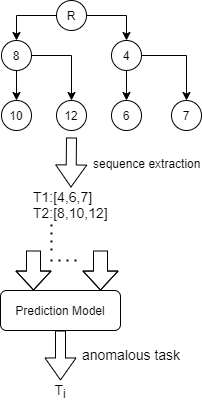}
        \caption{}
        \label{DeepTaskAPT Sequence creation}
    \end{subfigure}%
    
\caption{DeepTaskAPT cyber threat detection: (a) system architecture (b) sequence creation approach from a task tree   } \label{fig:1}
\end{figure*}

DeepTaskAPT addresses the same issue with a solution based on a task-tree based deep learning model (see Figure. \ref{DeepTaskAPT Sequence creation}). More clearly, instead of using a deep learning model directly, we demonstrate a task-based tree indicating process-oriented user behavior. We can find out anomalous user's tasks based on the relationship between operations. We rely on the process-oriented nature of operating systems and assume that prediction approaches can suffer a reliability loss if task-based relationships in the log are ignored. DeepTaskAPT can detect anomalous tasks based on a complete or partial sequence of operations in the task from host-based telemetry records captured on a single host or across an enterprise-size network. From this detector, our second goal is to train such models from telemetry data under weaker assumptions than previous work.

DeepTaskAPT, like other models, can be trained using only normal/benign data. Rather than using a multi-source LSTM model whose training requires telemetry describing both benign and malicious activity, it leverages an LSTM network to encode users' log templates and to predict the next action in the sequence.




\begin{table*}[h]
\caption{Log Entries From OpTC Datset}\label{Result}

\resizebox{\linewidth}{!}{%
\begin{tabular}{|l|l|l|l|l|l|l|l|l|l|l|l|c|}
\hline
id       & object  & action & pid  & ppid & actorid  & objectid & principal                                                                                   & file\_path         & image\_path & parent\_image\_path & timestamp                  & malicious \\ \hline
a390127d & FILE    & CREATE & 4    & 0    & 1f8b17b2 & 82ecf099 & NT AUTHORITY\textbackslash{}textbackslash\{\}\textbackslash{}textbackslash\{\}SYSTEM        & nan                & System      & nan                 & 2019-09-25 12:32:14.303000 & 0         \\ \hline
d4f73408 & PROCESS & START  & 1804 & 554  & 6600a6eb & d2bb8111 & NT AUTHORITY\textbackslash{}textbackslash\{\}\textbackslash{}textbackslash\{\}SYSTEM        & winlogbeat.yml.new & lwabeat.exe & nan                 & 2019-09-25 15:38:13.715000 & 1         \\ \hline
4288ccff & FILE    & READ   & 344  & 556  & a0731a85 & 7fe74abd & NT AUTHORITY\textbackslash{}textbackslash\{\}\textbackslash{}textbackslash\{\}LOCAL SERVICE & Security.evtx      & svchost.exe & nan                 & 2019-09-25 15:38:14.552000 & 0         \\ \hline
...      & ...     & ...    & ...  & ...  & ...      & ...      &                                                                                             & ...                & ...         & ...                 & ...                        & ...       \\ \hline
\end{tabular}
}
\label{Log Entries From OpTC Datset}
\end{table*}

\subsection{System architecture}
DeepTaskAPT architecture, as shown in Figure. \ref{fig:1}, consists of three components: Task-tree construction, building a baseline model, and anomaly detection.

\subsubsection{\textbf{Task tree construction} }
\label{sub:ConstructingProcesstree}

General-purpose operating systems start new processes in the context of another one that's already running: the former is deemed the \emph{child} process, and the latter, the \emph{parent} process. This filiation relationship establishes a hierarchy across all running computations on a system. Host-based sensors described in Section \ref{sec:introduction} report the start of any new process, and annotate it with an identifier of the parent process. Such process identifiers also tag any other event reported by the sensor (such as network transactions and file operations), augmenting the process tree with a set of leaf objects. We call this augmented tree the \emph{task tree}. We call a \emph{task} the union of a process and the events performed in its context, which correspond to its child nodes in the task tree and
a \emph{trace} is a sequence of events in a task ordered chronologically.

To create tasks for a user from a host-based telemetry stream, we rely on process identifiers associated with each record, and assume that process-start events are also tagged with the parent process identifier. The goal of task tree generation is to map the relationships between log events representing normal user behaviour and reveal malicious operations. 


DeepTaskAPT tree construction algorithm tackles all the use cases possible to generate a complete or partial task tree and splits the events into tasks that we store in a tree to find out the relationship between events. 
We use $\mathsf{CreateTaskTree}$ function in Algorithm. \ref{Algo: Tree Creation} to construct the task tree. It begins with the events in sorted order and each event is then added to the tree based on its filiation relationship such as child (process\_id, object\_id) and parent (parent\_process\_id, actor\_id) in OpTC data.
A detailed description of our tree construction process is given in Algorithm. \ref{Algo: Tree Creation} and a sample tree in Figure. \ref{DeepTaskAPT Sequence creation}. 
Each child of the default root R is treated as a \textit{task} and all events that are part of the task are treated as \textit{traces} under the task.





\begin{algorithm}[h]\footnotesize
\caption{Task Tree Construction}
\begin{algorithmic}[1]
\Procedure{CreateTaskTree}{$D$}\Comment{D: List of actions from Dataset}\\
\textbf{Output:} $tree$ \Comment{tree= tree of all actions from D}

\State $tree \gets$ new $Tree$
\State $tree.addnode(R)$
\State $count \gets 0$
\For{ $action$ in $D$}
    \State $nodeid \gets (action.pid,action.objectid)$
    \State	$parentid \gets (action.ppid, action.actorid)$
    \If{$nodeid$ not in $tree$}
        \If{$parentid$ not in $tree$}
        \State $tree.addnode(id=parentid, parent=R)$
        \EndIf
        \State $tree.addnode(id=nodeid, parent=parentid)$
    \Else
        \If{tree.nodeid.parent = R}
            \If{$parentid$ not in $tree$}
                \State $tree.addnode(id=parentid, parent=R)$
            \EndIf
			\State $tree.move(nodeid,parentid)$
        \Else
            \If{$tree.nodeid.parent \neq parentid$}
                \State $count \gets FlagNodes(tree, nodeid, count)$
                \If{$parentid$ not in $tree$}
                \State $tree.addnode(id=parentid, parent=R)$
                \EndIf
                \State $tree.addnode(id=nodeid, parent=parentid)$
            \Else
                \State$tree.nodeid.adddata(action)$
            \EndIf
        \EndIf
    \EndIf
\EndFor
\Return $tree$
\EndProcedure
\end{algorithmic}
\label{Algo: Tree Creation}
\end{algorithm}

\begin{algorithm}[h]\footnotesize
\caption{Flag Nodes}
\begin{algorithmic}[1]
\Procedure{FlagNodes}{$tree,id,count$}\Comment{tree:task tree, id: node to be flagged, count: \#flagged nodes}\\
\textbf{Output:} $count$ \Comment{count= new \#flagged nodes}

\State $node\gets tree.node(id)$
\State $newid \gets id + count$
\State $tree.addnode(id=newid)$
\State $count \gets  count +1$
\For{ $child$ in $node$}
    \State $returnid , count \gets FlagNodes(Tree,child.id,count)$
    \State $tree.movenode(returnid, newid)$
\EndFor
\State $tree.removenode(id)$

\Return $count$
\EndProcedure
\end{algorithmic}
\end{algorithm}






\begin{algorithm}  \footnotesize
\caption{Encoding keys in a Task}
\begin{algorithmic}[1]
\Procedure{Encode}{$K, n$}\Comment{K= actions in task n: \#unique actions}\\
\textbf{Output:} $r$ \Comment{r= encoded actions}
\State $r \gets \emptyset$ 
\State $flag\gets$ False
\For{ $action$ in $K$}
    \If{$action$ = $K.last$}
        \If{ $action.last \neq action$  and not  $flag $}
        \State $r.add(action)$
        \EndIf
    \Else
        \If{$action.last \neq action$  }
            \State $r.add(action)$
            \State $flag\gets$ False
        \Else
            \If{$action.last \neq action.next$ and not $flag$ }
            \State $r.add(action)$
            \Else
                \If{not $flag$}
                    \State $r.add(action + n)$
                    \State $flag\gets$ True
                \EndIf
            \EndIf
        \EndIf
    \EndIf
\EndFor
\Return $r$
\EndProcedure
\end{algorithmic}
\label{Algo:Encoding}
\end{algorithm}

\subsubsection{\textbf{Encoding Task sequence}}
\label{sec:encoding}

 To obtain a logical representation of a user's operations, we develop a log sequence encoder that handles contextual knowledge such as repetition of action in a log sequence. The purpose of the encoding algorithm is to ensure a meaningful sequence of actions when the repetitive occurrence of certain actions may fill the window. For instance, the File-Creation action occurs 1000 times consecutively. Instead of removing these action sequences completely, we encode them as a new key. 
 
 We use $\mathsf{Encode}$ function as described in the Algorithm. ~\ref{Algo:Encoding} to generate encoded keys in a task. $\mathsf{Encode}$ gives a new label to the events that occurred more than two times consecutively. Let a sequence be $\{a\quad b\quad b \quad b\quad b \quad b\quad c \}$. We relabelled the sequence as $\{a \quad b \quad b \quad b^{\prime} \quad c\}$. However, $\{a\quad b \quad b \quad c\}$ would remain as is. 
The encoding process occurs before the task sequence is passed to the anomaly detection models.

The encoding algorithm takes in a sequence of operations to encode and the total number of unique operations in a dataset (e.g. 32 in OpTC data). This ensures that the output sequence has a maximum of two duplicate items consecutively. We observed 61 encoded keys after the applying encoding function on 32 unique actions in OpTC data, and 8 encoded keys from 4 unique keys in LANL data.

\subsubsection{\textbf{Building baseline model}}

We interpret log entries from a user into trees or chronological sequences. They are all connected to form a heterogeneous forest. Each tree in the forest, corresponding to a task, is derived from a process based relationship (see Section. \ref{sub:ConstructingProcesstree}). DeepTaskAPT's baseline model includes all types of normal host-based events in a joint model. For baseline construction, we randomly choose users in the dataset with a good number of log entries involved in daily activities comprising all operations (log keys). The object-action pairs (OpTC) or eventIDs (LANL), as referred to \textit{log keys}, parsed from the task trees, are used to train the LSTM based anomaly detection model. The LSTM network is a classifier that is trained entirely on normal usage data, with no assumptions about malicious tactics or common malware categories.



\subsubsection{\textbf{Detection and validation}}
DeepTaskAPT adopts the baseline anomaly detection model to analyze a task to be benign or malicious. Assuming that the newly arrived user events are parsed into a task (or subtask) and then the sequence of task actions, DeepTaskAPT determines if the incoming task is malicious. A task is labelled as malicious if one or more entries in the task are predicted to be malicious. However, given the expert input, the observed anomaly could be a false positive. The model can be updated to integrate and conform to the new trend. In order to validate DeepTaskAPT performance, we use two types of public datasets: synthetic (DARPA OpTC), real-life (LANL); and compare with similar existing models such as Linear Regression, DeepLog, Random Forest. We evaluate the results with testing data from the same or different users against the baseline model.    







\subsection{Threat model}
In general, host based system logs collected from network sensors are considered secure and private in a large enterprise setting. However, APT attacks, typically triggered by insider employees, may perform malicious activities such as installing malicious software, data leaks, scanning the system for vulnerability, compromising other hosts for escalating privileges using the APT actor's valid credentials. APT can be modeled through three main approaches: asset-centric approach focuses on individual assets that have value to the attacker, a system-centric approach finds out vulnerability in the overall system software, and a data-centric approach prevents data leakage \cite{APTmodel}. DeepTaskAPT concentrates on identifying the approaches and a model that can detect malicious activities and hence reduce APT risk.

\subsection{Dataset for Cyber threat detection}
We validate our anomaly detector using DARPA's Operationally Transparent Computing Cyber (OpTC) dataset \cite{OpTC2020}. This dataset is the most detailed public dataset that includes host-based telemetry records. Indeed, this sort of dataset is typically gathered privately, either as part of an enterprise's own cybersecurity operations, or by running professional security services. While free sensor and data centralization software has been available for a few years \cite{Rodriguez2020}, building such a dataset from scratch is a difficult endeavor, notably involving privacy issues if real users are involved in the data collection. Thus, the public first dataset recognizable as host-based telemetry was heavily crippled for anonymization, discarding even process filiation information \cite{Kent2015}. Its authors followed it up with an improved dataset \cite{Turcotte2018}, whose host-based component adopted a richer variable schema, much closer to comparable private datasets. However, this stream only described normal activity on an IT network: unlike its predecessor, no red team operation went on during its collection, which precludes the validation of anomaly detectors as proxies to malicious activity detectors. 
The OpTC dataset provides three types of red team engagements that mirror such modern tactics, and its baseline activity, while still generated through simulators, echoes the structure and complexity of the private datasets the authors have used in their own cyber defense research work. As mentioned early, the LANL Unified Host and Network dataset \cite{Turcotte2018} is the only available public dataset that is somewhat close to the OpTC dataset. It captures LANL's network and host operations over the span of 90 days. Similar to the OpTC dataset, this dataset reports detailed process information required for DeepTaskAPT's task-tree generation. It is worth noting that we couldn't use the LANL 2015 dataset \cite{Kent2015} due to the missing filiation relationship in the system process.




\begin{table}[!htb]
\caption{Experiment datasets}
\begin{adjustbox}{width=\columnwidth,center}
\label{Table:SetupDataset}
\centering
\begin{tabular}{|c|c|c|c|}
\hline
Log Dataset                                                         & \#Training Data                                                                              & \#Test Data                                                                                                                               & \#encoded keys \\ \hline
\begin{tabular}[c]{@{}c@{}}OpTC \\ (same user)\end{tabular}          & \begin{tabular}[c]{@{}c@{}} 25000 (task)\\ 1015441 (trace)\end{tabular} & \begin{tabular}[c]{@{}c@{}}\#labeled 1 (task)\\ \#unlabeled 17898 (task)\\ \#labeled 14142 (traces)\\ \#unlabeled 1355221 (trace) \end{tabular} & 61        \\ \hline

\begin{tabular}[c]{@{}c@{}}OpTC\\  (different user)\end{tabular} & \begin{tabular}[c]{@{}c@{}} 42898 (task)\\   2975266(trace) \end{tabular} & \begin{tabular}[c]{@{}c@{}}\#labeled 36 (task)\\ \#unlabeled 8296 (task)\\ \#labeled  53461 (trace)\\ \#unlabeled  471596 (trace)\end{tabular}   & 61             \\ \hline

\begin{tabular}[c]{@{}c@{}}LANL \\ (different user)\end{tabular}     & \begin{tabular}[c]{@{}c@{}} 5 (task)\\ 41828 (trace) \end{tabular}       & 

\begin{tabular}[c]{@{}c@{}}\#unlabeled 14 (task)\\ \#unlabeled 31138 (trace) \end{tabular}                                                   & 8              \\ \hline
\end{tabular}
\end{adjustbox}
\end{table}

\section{Experiment}

We conduct our experiment on a workstation with an Intel Core i7-10750H running at 2.6 GHz, 128G RAM and 6 cores. We use spark 3.0.2, python libraries for data extraction, tree generation, training and tuning deep learning model.

\subsection{Experimental Dataset and methods }

OpTC/ecar logs were produced by several different threads or concurrently running tasks. As part of pre-processing, DeepTaskAPT applied a Task-based tree construction algorithm (see Section \ref{sub:ConstructingProcesstree}) followed by an encoding function (see Section \ref{sec:encoding}) to identify individual tasks of users to train and test the model. As mentioned earlier, the OpTC dataset is highly imbalanced in the number of labelled or malicious events. For example, we identified 6 malicious tasks out of 27099 tasks for user ID user0352, 1 malicious task out of 42899 tasks for user0201. We assume a task is malicious if at least one of its traces is labelled as malicious. Class imbalance is a well-known issue in the area of machine learning or deep learning.This is obvious in the OpTC dataset, with labelled items being approximately 0.0016\% of the whole dataset from 27 users. The disparity in the amount of benign and malicious events makes it difficult to train and test models as the model performance degrades, especially for the minority classes. DeepTaskAPT presents two solutions to tackle this problem: 1) training the model with only benign or unlabelled data, 2) translate the events in the task to window-size traces for the experiment. For instance, we created 14142 malicious traces from 1 malicious task for OpTC user0201, 41131 benign traces from 1 task for LANL user024735.

We collect 6 days (90 days) of users' tasks from OpTC data (LANL data) and transformed them to sequences of events to learn the representations of ordinary users. The model that emerges will then be used to identify labelled/malicious tasks or actions performed by the same or other users. 
For the OpTC dataset, we present test results from the model trained by one user's benign data and validated by the benign and malicious data from 13 users. In contrast, the LANL dataset does not have any labelled malicious events. Moreover, it lacks the richness of contextual facts that hinder fine-grained feature engineering. For instance, only 4 operations (out of 20 in the dataset) have filiation information in the LANL dataset, in contrast to 32 (out of 32 in the dataset) in the OpTC dataset. We use this dataset only to validate the performance of our baseline model with the accuracy metric.

Events Log for the user has been extracted from (OpTC data) to parse to a 'object-action', 'pid-ppid', 'actorid-objectid' values vector to generate a task tree.  
The LANL dataset has 4 types of events with filiation information (login and process events only). Tasks trees for each user in the LANL dataset is created with 'EventID', 'ProcessID-ParentProcessID', 'ProcessName-ParentProcessName' values vector. A detail description of task tree generation is given in the section. \ref{sub:ConstructingProcesstree}.
Due to resource constraints, for the OpTC dataset DeepTaskAPT considers only the first few occurrences (e.g. 300, 1000, 1500) rather than training all of the traces/sequences in the task.

Experimental data from LANL only includes trace based evaluation. Due to a data shortage, we train the model with 41828 traces from five random users and test the model on 31138 traces from 14 users. A detailed description of experiment data is given in Table. \ref{Table:SetupDataset}.










\subsection{Anomaly detection}


DeepTaskAPT trains the model as a multi-classifier over recent user task operations where input is a history of recent task based actions/traces (keys), and the output is a probability distribution over the number of classes so that it can predict the probability of the next operation in a sequence of operations. Suppose we have a task resulting from benign execution parsed into a sequence of actions $\{a_i, a_{i+1},a_{i+2}…, a_n\}$. Given a window size ($w=15$ in our case), we create an input sequence and output level $\{a_j, a_{j+1}…a_{15} \rightarrow a_k\}$. In the detection phase, we send a window from a task to the model as its input. The output will be the probability distribution of each candidate to be the next action. If an action is among the top $t$ candidates, DeepTaskAPT treats it as normal, otherwise malicious. This is similar to a traditional $N$-gram model where $N$ is the window size. Fig: \ref{DeepTaskAPT Architecture} embodied our design. 


In the training phase, the model decides on proper weight allotments to produce the desired output in the final output of the LSTM sequence. Each input-output pair updates these weights incrementally throughout the training phase by minimising losses (categorical cross-entropy loss) thru gradient descent where an input is a window $w$ with $t$ operation keys, and an output is the action key value that follows $w$. In the detection phase, the input layer is the encoded one hot vector of the $t$ potential log keys from $\mathsf{G}$. An output $k_t$ is predicted from the input window $w = (k_0, k_1, \dots, k_{(t-1}))$ using a layer of LSTM blocks. The output layer actually converts the hidden state into a distributed probability function $Pr (k_t = p_i|w)$ s.t. $p_i \in \mathsf{G}$.  \\

\noindent \textbf{{Metrics:}} For evaluation, we performed both task based and trace based prediction. In the \textit{task-based} prediction, the anomaly detection system determines True Positive (TP) and False Positive (FP) based on the first occurrence of miss-classification. While \textit{trace-based} prediction evaluates all the traces in the test dataset.

\begin{itemize}
    \item TP: if malicious labelled is not predicted correctly
    \item FP: if benign labelled is not predicted correctly

\end{itemize}


\begin{table}[h]
\begin{tabular}{|l|c|}
\hline
\textbf{Metric}           & \textbf{Computation detail}                          \\ \hline
Sensitivity/Recall        & TP / (TP + FN)                                       \\ \hline
False Positive Rate (FPR) & FP / (FP + TN)                                       \\ \hline
Specificity               & TN / (FP + TN)                                       \\ \hline
Accuracy                  & (TP + TN)/ (TP + TN + FP + FN)                               \\ \hline
Precision                 & TP / (TP + FP)                                       \\ \hline
GMean                     & $\sqrt{\mathrm{Sensitivity} * \mathrm{Specificity}}$ \\ \hline
\end{tabular}
\end{table}

\subsection{Model Parameters}
DeepTaskAPT's baseline model construction includes generating a complete task tree for the target user followed by training the LSTM model. We use the gradient descent with decaying learning rate for the error calculation, categorical cross-entropy as a loss function. DeepTaskAPT uses window size $w=15$, number of layers $L= 2$, the number of memory units per block $\alpha = 64$, batch size $B =2048$, number of epochs $ \epsilon = $153 to 250. \#candidate is the number of options DeepTaskAPT is able to compare to for each trace/task, the higher numbers allow for lower FPR but also lower recall. The number of predicted \#candidates (3 to 19) has been adjusted based on the performance requirement.  
The random forest model used in the experiment employs 50 estimators with no maximum depth. All other settings are left at default.

\section{Results and discussion}
\label{sec:results}

\begin{figure*}[htb]
\minipage{0.25\textwidth}
  \includegraphics[width=\linewidth]{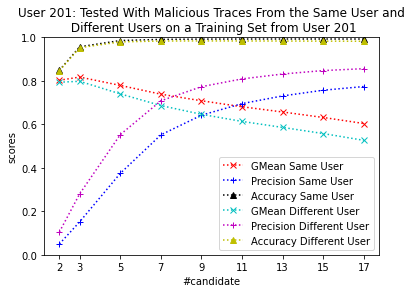}
\endminipage\hfill
\minipage{0.25\textwidth}
  \includegraphics[width=\linewidth]{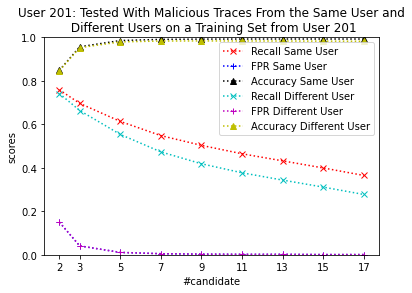}
\endminipage\hfill
\minipage{0.25\textwidth}%
  \includegraphics[width=\linewidth]{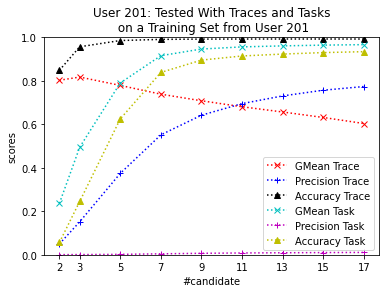}
\endminipage\hfill
\minipage{0.25\textwidth}%
  \includegraphics[width=\linewidth]{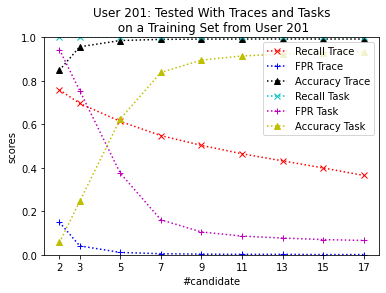}
\endminipage
\caption{Performance evaluation on OpTC Dataset user0201 (same user)}
\label{fig201}
\end{figure*}

\begin{figure*}[!htb]
\minipage{0.25\textwidth}
  \includegraphics[width=\linewidth]{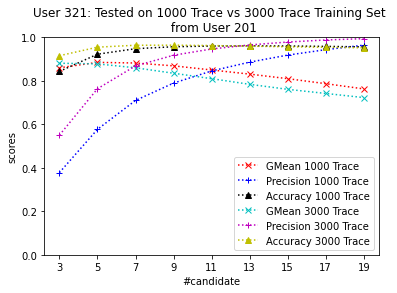}
\endminipage\hfill
\minipage{0.25\textwidth}
  \includegraphics[width=\linewidth]{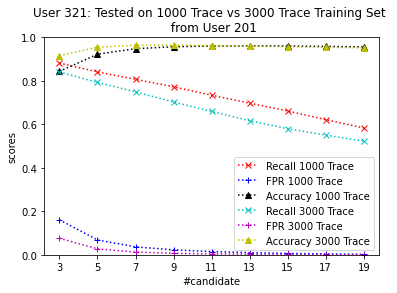}
\endminipage\hfill
\minipage{0.25\textwidth}%
  \includegraphics[width=\linewidth]{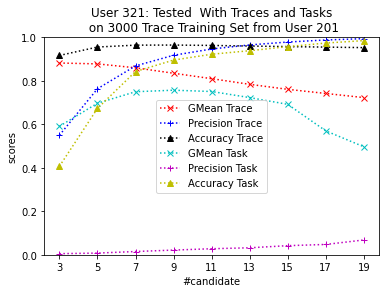}
\endminipage\hfill
\minipage{0.25\textwidth}%
  \includegraphics[width=\linewidth]{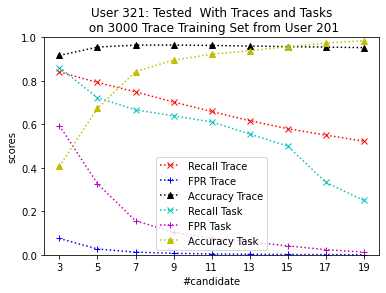}
\endminipage
\caption{Performance evaluation on OpTC Dataset training: user0201 testing: user0321 window size: 15}
\label{fig321}
\end{figure*}

We evaluate the performance of DeepTaskAPT based on DeepLog \cite{Du2017} and typical machine learning algorithms such as Random Forest, Linear Regression etc. In order to compare our approach with other similar approaches, we build a comprehensive train/test dataset (see Table. \ref{Table:SetupDataset}) from the OpTC and LANL datasets. According to our findings, DeepTaskAPT outperforms existing related approaches in log based anomaly detection. Besides, the efficacy of task based tree and the feature set's diversity in the OpTC dataset aids in the analysis of activity sequences by detection models. The metrics that we employ to compare different approaches are accuracy, FPR, recall and G-Mean.

The performance of DeepTaskAPT on the OpTC (same user) dataset can be seen in Figure. \ref{fig201} and Table. \ref{Table: Task entries}. 
When the model is tested only with malicious traces from the same user, the model achieves better classification performance than when tested against malicious traces from all users. 
Evaluating the performance of using traces and tasks from the same user, the model shows that traces have a higher accuracy score and a lower FPR than using tasks. However, the model with tasks is able to achieve a recall score of 1.    

\begin{table}[!htb]
\caption{Anomaly Detection results for DeepTaskAPT with Tasks on  OpTC Dataset with malicious Tasks from Same User and All Users}
\label{Table: Task entries}
\centering
\begin{tabular}{|c|c|c|}
\hline
Test User                                                                        & Same User & All User \\ \hline
\begin{tabular}[c]{@{}c@{}}\# of  detected log\\  entries/ \# total\end{tabular} & 13/13     & 34/36    \\ \hline
\end{tabular}
\end{table}

Table. \ref{Table: Task entries} shows anomaly detection results for DeepTaskAPT when using tasks. When only testing with anomalous tasks from the same user (user0201), the model is able to detect all of the anomalous tasks 13 out of 13. With anomalous tasks from all users in the dataset, 34 out of 36 anomalous tasks are detected.

The results for user0321 and user0205 using a DeepTaskAPT model trained on user0201 data can be seen in Figure. \ref{fig321} for user0321.  The results for both users show that with an increased number of training traces, the model's accuracy improves and has a lower FPR, but this also results in reduced recall performance. For user0321, comparing the model's performance when using traces versus tasks shows that the performance for traces is better for all metrics except for accuracy at 15 candidates or greater. While for user0205, the relative performance is dependent on the number of candidates as tasks have higher recall below 11 candidates and higher accuracy above 17 candidates.

The performance of a DeepTaskAPT model trained on user0201 applied to two different users, user0321 and user0205, shows that a single model can be applied to different users within a dataset without significant changes to performance. The model's anomaly detection performance for both users is similar, with user0205 achieving a higher accuracy but having a near-identical FPR and recall performance. This result implies that it would be possible to generalize a DeepTaskAPT model trained on one user in a dataset and apply that model to different users within the same dataset as the individual user tested does not heavily impact the model's performance.


Comparing the performance of DeepTaskAPT to DeepLog on the OpTC (same user) dataset (see Table. \ref{Table:OpTC Accuracy}), DeepTaskAPT results in better performance in all metrics regardless of the number of candidates chosen. 
When both models are applied to the OpTC (different user) dataset, DeepTaskAPT achieves a lower FPR (0.0529 vs. 0.1488) and higher accuracy (0.9363 vs 0.8204) as well as having a much higher recall score (0.882 vs. 0.5834). 


\begin{table}[h]
\centering
\caption{Anomaly Detection Performance for Different Models on OpTC Dataset using traces with Models Trained on user0201 and tested on user0201 and user0205. \#candidate=5 for RF, DeepLog and DeepTaskAPT.}

\begin{tabular}{|c|c|c|c|c|}
\hline
\multirow{2}{*}{Method} & \multicolumn{2}{c|}{user0201} & \multicolumn{2}{c|}{user0205} \\ \cline{2-5} 
                        & Accuracy       & FP Rate      & Accuracy       & FP Rate      \\ \hline
DeepTaskAPT                 & 0.9854         & 0.011        & 0.9363         & 0.0529       \\ \hline
DeepLog                 & 0.8354         & 0.161        & 0.8204          & 0.1488       \\ \hline
RF (tree-processed)               & 0.9052         & 0.0833       & 0.8739         & 0.0921       \\ \hline
RF (raw)           & 0.8088          & 0.1831       & 0.7967         & 0.1800       \\ \hline
LR (tree-processed)               & 0.1139         & 0.9145       & 0.1635         & 0.9239       \\ \hline
LR (raw)           & 0.0858         & 0.9370       & 0.1128         & 0.9566       \\ \hline
\end{tabular}

\label{Table:OpTC Accuracy}
\end{table}

\begin{table}[h]
\centering
\caption{Anomaly Detection Performance for Different Models on OpTC Dataset using traces with Models Trained on user0201 and tested on user0201  and user0205. \#candidate=2 for DeepLog (user0201) and DeepTaskAPT(user0201) \#candidate=3 for RF, DeepLog(user0205) and DeepTaskAPT(user0205). }

\begin{tabular}{|c|c|c|}
\hline
\multirow{2}{*}{Method} & user0201 & user0205 \\ \cline{2-3} 
                        & Recall   & Recall   \\ \hline
DeepTaskAPT                 & 0.7587   & 0.882   \\ \hline
DeepLog                 & 0.7202   & 0.5834   \\ \hline
RF (tree-processed)               & 0.6784   & 0.6784   \\ \hline
RF (raw)           & 0.6132   & 0.6132   \\ \hline
LR (tree-processed)               & 0.9339   & 0.9344   \\ \hline
LR (raw)           & 0.9057   & 0.9466   \\ \hline
\end{tabular}

\label{Table:OpTC Recall}
\end{table}

DeepTaskAPT's task tree construction is able to process data for different prediction models such as random forests (RF) and linear regression (LR) models. \textit{Task-tree}  processed and \textit{raw} data are tested with RF and LR models to evaluate the performance impact of using the processed data. Table. \ref{Table:OpTC Accuracy} and Table. \ref{Table:OpTC Recall} show the performance of the different methods on the OpTC dataset. DeepTaskAPT achieves the highest performance out of all of the approaches tested, with the highest accuracy (0.9854) and the lowest false positive rate (0.011), as well as one of the highest recall scores (0.882). The models that use tree processed data from DeepTaskAPT's task tree generation deliver better performance than the same type of model using raw data. This behaviour is seen in both the random forest and linear regression models. This performance improvement results in the random forest model with tree processed data outperforming DeepLog in all metrics except for user0201 recall performance. In comparison, the random forest model with raw data performs worse than DeepLog in all metrics except for recall performance for user0205. 
The linear regression model with tree processed data outperforms in all metrics for both users except for recall preference on user0205. Although the recall appears to be incredibly good, LR is not the winner of the experiment due to its high FPR and the low accuracy.

\begin{table}[!htb]
\caption{Number of FPs and FNs on OpTC Dataset using traces with Models Trained on user0201 and Tested on user0205 \#candidate=5 for RF, DeepLog and DeepTaskAPT.}
\begin{adjustbox}{width=\columnwidth,center}
\label{Table: OpTC FPFN}
\begin{tabular}{|c|c|c|c|c|c|c|c|}
\hline
Method                                                                     & DeepTaskAPT-Trace                                           & DeepTaskAPT-Task & DeepLog                                                 & RF (tree-processed)                                     & RF (raw)                                                & LR (tree-processed)                                      & LR (raw)                                                 \\ \hline
\begin{tabular}[c]{@{}c@{}}false positive (FP)\\ /\#Unlabeled\end{tabular} & \begin{tabular}[c]{@{}c@{}}24971\\ /471596\end{tabular} & 3274/8296    & \begin{tabular}[c]{@{}c@{}}68622\\ /461201\end{tabular} & \begin{tabular}[c]{@{}c@{}}43426\\ /471545\end{tabular} & \begin{tabular}[c]{@{}c@{}}90004\\ /500000\end{tabular} & \begin{tabular}[c]{@{}c@{}}435640\\ /471545\end{tabular} & \begin{tabular}[c]{@{}c@{}}478314/\\ 500000\end{tabular} \\ \hline
\begin{tabular}[c]{@{}c@{}}false negative (FN)\\ /\#Labeled\end{tabular}   & \begin{tabular}[c]{@{}c@{}}8493\\ /53461\end{tabular}   & 6/36         & \begin{tabular}[c]{@{}c@{}}21573\\ /41048\end{tabular}  & \begin{tabular}[c]{@{}c@{}}22761\\ /53461\end{tabular}  & \begin{tabular}[c]{@{}c@{}}20129\\ /41661\end{tabular}  & \begin{tabular}[c]{@{}c@{}}3505\\ /53461\end{tabular}    & \begin{tabular}[c]{@{}c@{}}2225\\ /41661\end{tabular}    \\ \hline
\end{tabular}
\end{adjustbox}
\end{table}

Table. \ref{Table: OpTC FPFN} shows the number of false-positive and false-negative results for each method.
The linear regression (raw) method achieves the lowest number of false-positive results but also generates the highest number of false-negative results showing the unbalanced performance of this method. DeepTaskAPT has the best overall performance generating the smallest percentage of false results. 

\begin{table}[!htb]
\caption{Anomaly Detection results for Different Models on OpTC Dataset using traces with  Models  Trained  on  user0201 and Tested on user0205 \#candidate=3 for RF, DeepLog and DeepTaskAPT.}
\begin{adjustbox}{width=\columnwidth,center}
\label{Table: OpTC entries}
\begin{tabular}{|c|c|c|c|c|c|c|}
\hline
Method                                                                           & DeepTaskAPT     & DeepLog     & RF (tree-processed)   & RF (raw) & LR (tree-processed)   & LR (raw) \\ \hline
\begin{tabular}[c]{@{}c@{}}\# of  detected log \\ entries/ \#total\end{tabular} & 47150/53461 & 23946/41048 & 45367/53461 & 34078/41661   & 49956/53461 & 39436/41661   \\ \hline
\end{tabular}
\end{adjustbox}
\end{table}

Table. \ref{Table: OpTC entries} shows the anomaly detection results for each method using traces. Out of all of the methods, linear regression (raw) followed by linear regression (tree-processed) achieves the highest percentage of anomalous events detected, but these methods also had the highest FPR out of all methods. For methods with an acceptable FPR,  
DeepTaskAPT results in the highest percentage of anomalous events detected, followed by the random forest models, and lastly, DeepLog. DeepLog exhibits better performance when tested and trained on the same user.

\begin{table}[h]
\centering
\caption{Classification Accuracy on LANL Dataset with DeepTaskAPT and DeepLog Models. Both models trained on 5 users, models tested on 14 users. \#candidate=3 for DeepLog and DeepTaskAPT}

\begin{tabular}{|c|c|}
\hline
Method  & Accuracy \\ \hline
DeepTaskAPT                 & 0.983    \\ \hline
DeepLog                & 0.883    \\ \hline
\end{tabular}

\label{Table:LANL}
\end{table}

Using the LANL dataset, accuracy is measured based on if the predicted value from the model matches the following value in the sequence as LANL is an unlabeled dataset.  Table. \ref{Table:LANL} shows that DeepTaskAPT achieves a higher prediction accuracy than DeepLog (0.983 vs 0.883) when tested on the same users. DeepTaskAPT can outperform DeepLog on both the OpTC dataset and the LANL dataset.


\section{Conclusion}
\label{sec:conclusion}



Task trees package sequential details of log operations that are chronologically distant yet semantically close. This enables DeepTaskAPT a useful model to detect APT attacks as APT also reflects this characteristic. Task tree based sequence crea tion is thus an important step in creating efficient event representations for LSTM-based sequence classification. While DeepTaskAPT effectively detects anomalous behaviour in the OpTC dataset with high accuracy and a low FPR, we demonstrate that the task tree generation method can improve the performance of other prediction methods.


Training deep learning models is a big challenge given computations and memory limitations. We observed that training the model with more data and the number of epochs (model parameter) improves the results. For example, training the model with the first 1500 traces from the tasks yields better results than the trained model with the first 300 traces. It is worth noting that due to resource (RAM) limitations, we were unable to train the model with full user data or epoch values greater than 157 in certain instances. In an enterprise setting, distinct baseline models for various types of users may be developed. Alternatively, a baseline model may be constructed using data from various groups of users. Furthermore, a hierarchical architecture for anomaly detection can be developed. For example, after initial detection, a second level vector with additional parameters such as frequency values can be produced. If any actions or parameters are predicted to be malicious, the new task will be labelled as malicious. Since the model can be retrained/updated with new users' data, these plans will be carried out in the future.     

\section*{Acknowledgement}
We thank Dr. Benoit Hamelin from Communications Security Establishment Canada for his valuable and constructive suggestions. His willingness to share malicious labelled data with technical details so generously has been very much appreciated.



\end{document}